%
\catcode`@=11
\global\newcount\secno
\global\newcount\subsecno
\global\newcount\subsubsecno
\global\newcount\equationno
\global\newcount\refno
\global\newcount\footnoteno
\global\newcount\@no
\secno=0\subsecno=0\subsubsecno=0\equationno=0\refno=0\footnoteno=0
\let\\=\cr
\def\@draftleft#1{}
\def\@draftright#1{}
\overfullrule=0pt
\def\draft{\def\@draftleft##1{\leavevmode\vadjust{\smash{%
\raise3pt\llap{\eighttt\string##1~~}}}}%
\def\@draftright##1{\rlap{\eighttt~~\string##1}}%
\def\date##1{\leftline{\number\month/\number\day/\number\year\
\the\time}\bigskip}\overfullrule=5pt\normalbaselineskip=18pt
\normalbaselines}
\def\@the#1{\ifnum\the#1>0\relax\the#1\else\ifnum\the#1<0\relax
\@no=-\the#1\advance\@no'100\char\@no\fi\fi}
\def\@advance#1{\ifnum\the#1<0\global\advance#1 -1\relax
\else\global\advance#1 1\relax\fi}
\def\nsec#1\par{\bigskip\allowbreak\bigskip\@advance\secno
\subsecno=0\subsubsecno=0\equationno=0
\vbox{\secfont\noindent
\@the\secno. #1\medskip}\nobreak\noindent\ignorespaces}
\def\secadvance{\@advance\secno}
\def\sec#1#2\par{\bigskip\allowbreak\bigskip
\subsecno=0\subsubsecno=0\equationno=0
\if*#1\vbox{\secfont\noindent\ignorespaces#2\medskip}%
\else
\secno=#1
\vbox{\secfont\noindent\@the\secno. #2\medskip}\fi
\nobreak\noindent\ignorespaces}
\def\seclab#1{\xdef#1{\@the\secno}\@draftleft#1}
\def\nsubsec#1\par{\bigskip\@advance\subsecno
\subsubsecno=0\equationno=0
\vbox{\subsecfont\noindent\@the\secno.\@the\subsecno. #1\medskip}%
\nobreak\noindent\ignorespaces}
\def\subsecadvance{\@advance\subsecno}
\def\subsec#1#2\par{\bigskip
\subsubsecno=0\equationno=0
\if*#1\vbox{\subsecfont\noindent\ignorespaces#2\medskip}%
\else
\subsecno=#1
\vbox{\subsecfon\noindentt\@the\secno.\@the\subsecno. #2\medskip}\fi
\nobreak\noindent\ignorespaces}
\def\subseclab#1{\xdef#1{\@the\secno.\@the\subsecno}%
\@draftleft#1}
\def\nsubsubsec#1\par{\medskip\@advance\subsubsecno
\vbox{\subsubsecfont\noindent
\@the\secno.\@the\subsecno.\@the\subsubsecno. #1\medskip}%
\nobreak\noindent\ignorespaces}
\def\subsubsecadvance{\@advance\subsubsecno}
\def\subsubsec#1#2\par{\medskip
\if*#1\vbox{\subsubsecfont\noindent\ignorespaces#2\medskip}%
\else
\subsubsecno=#1
\vbox{\subsubsecfont\noindent
\@the\secno.\@the\subsecno.\@the\subsubsecno. #2\medskip}\fi
\nobreak\noindent\ignorespaces}
\def\subsubseclab#1{\xdef#1{\@the\secno.\@the\subsecno.\@the\subsubsecno}%
\@draftleft#1}
\def\eqlabel#1{\@advance\equationno
\ifnum\secno=0\xdef#1{\the\equationno}\else
\ifnum\subsecno=0\xdef#1{\@the\secno.\the\equationno}\else
\xdef#1{\@the\secno.\@the\subsecno.\the\equationno}\fi\fi
\eqno({\eqnofont #1})\@draftright#1}
\def\lnlabel#1{\global\advance\equationno1
\ifnum\secno=0\xdef#1{\the\equationno}\else
\ifnum\subsecno=0\xdef#1{\@the\secno.\@the\equationno}\else
\xdef#1{\@the\secno.\@the\subsecno.\the\equationno}\fi\fi
&({\eqnofont #1})\@draftright#1}
\def\eqadvance#1{\global\advance\equationno1
\ifnum\secno=0\xdef#1{\the\equationno}\else
\ifnum\subsecno=0\xdef#1{\@the\secno.\the\equationno}\else
\xdef#1{\@the\secno.\@the\subsecno.\the\equationno}\fi\fi}
\def\eqlabelno(#1#2){\eqno({\eqnofont #1#2})\@draftright#1}
\def\lnlabelno(#1#2){&({\eqnofont #1#2})\@draftright#1}
\newwrite\rfile
\def\nref#1#2{\global\advance\refno1\xdef#1{\the\refno}%
\immediate\write
\rfile{\noexpand\item{#1.}\noexpand\@draftleft\noexpand#1%
#2}}
\def\sref#1#2{\immediate\write
\rfile{\noexpand\item{#1.}\noexpand\@draftleft\noexpand#1%
#2}}
\def\refs#1#2 {\if.#2#2$^{\rm#1}$\spacefactor=\sfcode`.{}\space
\else\if,#2#2$^{\rm#1}$\spacefactor=\sfcode`,{}\space
\else\if;#2#2$^{\rm#1}$\spacefactor=\sfcode`;{}\space
\else\if:#2#2$^{\rm#1}$\spacefactor=\sfcode`:{}\space
\else\if?#2#2$^{\rm#1}$\spacefactor=\sfcode`?{}\space
\else\if!#2#2$^{\rm#1}$\spacefactor=\sfcode`!{}\space
\else
$^{\rm#1}$#2\space\fi\fi\fi\fi\fi\fi}
\def\Refs#1{$\rm#1$}
\def\reportno#1{\line{\hfil\vbox{\halign{\strut##\hfil\cr#1\crcr}}}}
\def\Title#1{\vskip3\bigskipamount\line{\titlefont
\hfil\vbox{\halign{\strut\hfil##\hfil\cr#1\crcr}}\hfil}%
\vskip2\bigskipamount}
\def\author#1{\centerline{\authorfont#1}\medskip}
\def\address#1{\centerline{\vbox{\halign
{\strut\hfil\addressfont##\hfil\cr#1\crcr}}}%
\bigskip}
\def\abstract#1{{\narrower\abstractfont\null\bigskip\noindent\ignorespaces
#1\bigskip}}
\def\date#1{\leftline{#1}\bigskip}
\def\Tr{\mathop{\rm Tr}\nolimits}
\immediate\openout\rfile=refs.tmp
\font\seventeenrm=cmr17 \font\fourteenrm=cmr10 scaled 1440
\font\twelverm=cmr12  \font\eightrm=cmr8  \font\sixrm=cmr6
\font\seventeeni=cmmi10 scaled 1728 \font\fourteeni=cmmi10 scaled 1440
\font\twelvei=cmmi12  \font\eighti=cmmi8  \font\sixi=cmmi6
\font\seventeensy=cmsy10 scaled 1728 \font\fourteensy=cmsy10 scaled 1440
\font\twelvesy=cmsy10 scaled 1200 \font\eightsy=cmsy8 \font\sixsy=cmsy6
\font\seventeenbf=cmbx10 scaled 1728 \font\fourteenbf=cmbx10 scaled 1440
\font\twelvebf=cmbx12 \font\eightbf=cmbx8 \font\sixbf=cmbx6
\font\seventeentt=cmtt10 scaled 1728 \font\fourteentt=cmtt10 scaled 1440
\font\twelvett=cmtt12 \font\eighttt=cmtt8
\font\seventeenit=cmti10 scaled 1728 \font\fourteenit=cmti10 scaled 1440
\font\twelveit=cmti12 \font\eightit=cmti8
\font\seventeensl=cmsl10 scaled 1728 \font\fourteensl=cmsl10 scaled 1440
\font\twelvesl=cmsl12 \font\eightsl=cmsl8
\font\seventeenex=cmex10 scaled 1728 \font\fourteenex=cmex10 scaled 1440
\font\twelveex=cmex10 scaled 1200
\def\tenpoint{\def\rm{\fam0\tenrm}%
\textfont0=\tenrm\scriptfont0=\sevenrm\scriptscriptfont0=\fiverm
\textfont1=\teni\scriptfont1=\seveni\scriptscriptfont1=\fivei
\textfont2=\tensy\scriptfont2=\sevensy\scriptscriptfont2=\fivesy
\textfont3=\tenex\scriptfont3=\tenex\scriptscriptfont3=\tenex
\textfont\itfam=\tenit\def\it{\fam\itfam\tenit}%
\textfont\slfam=\tensl\def\sl{\fam\slfam\tensl}%
\textfont\ttfam=\tentt\def\tt{\fam\ttfam\tentt}%
\textfont\bffam=\tenbf\scriptfont\bffam=\sevenbf
\scriptscriptfont\bffam=\fivebf\def\bf{\fam\bffam\tenbf}%
\def\small{\eightpoint}%
\def\large{\twelvepoint}%
\def\normalbaselines{\lineskip\normallineskip
  \baselineskip\normalbaselineskip \lineskiplimit\normallineskiplimit}
\setbox\strutbox=\hbox{\vrule height8.5pt depth3.5pt width0pt}%
\normalbaselines\rm}
\def\twelvepoint{\def\rm{\fam0\twelverm}%
\textfont0=\twelverm\scriptfont0=\eightrm\scriptscriptfont0=\sixrm
\textfont1=\twelvei\scriptfont1=\eighti\scriptscriptfont1=\sixi
\textfont2=\twelvesy\scriptfont2=\eightsy\scriptscriptfont2=\sixsy
\textfont3=\twelveex\scriptfont3=\twelveex\scriptscriptfont3=\twelveex
\textfont\itfam=\twelveit\def\it{\fam\itfam\twelveit}%
\textfont\slfam=\twelvesl\def\sl{\fam\slfam\twelvesl}%
\textfont\ttfam=\twelvett\def\tt{\fam\ttfam\twelvett}%
\textfont\bffam=\twelvebf\scriptfont\bffam=\eightbf
\scriptscriptfont\bffam=\sixbf\def\bf{\fam\bffam\twelvebf}%
\def\small{\tenpoint}%
\def\large{\fourteenpoint}%
\def\normalbaselines{\lineskip1.2\normallineskip
  \baselineskip1.2\normalbaselineskip \lineskiplimit1.2\normallineskiplimit}
\setbox\strutbox=\hbox{\vrule height10.2pt depth4.2pt width0pt}%
\normalbaselines\rm}
\def\fourteenpoint{\def\rm{\fam0\fourteenrm}%
\textfont0=\fourteenrm\scriptfont0=\tenrm\scriptscriptfont0=\sevenrm
\textfont1=\fourteeni\scriptfont1=\teni\scriptscriptfont1=\seveni
\textfont2=\fourteensy\scriptfont2=\tensy\scriptscriptfont2=\sevensy
\textfont3=\fourteenex\scriptfont3=\fourteenex\scriptscriptfont3=\fourteenex
\textfont\itfam=\fourteenit\def\it{\fam\itfam\fourteenit}%
\textfont\slfam=\fourteensl\def\sl{\fam\slfam\fourteensl}%
\textfont\ttfam=\fourteentt\def\tt{\fam\ttfam\fourteentt}%
\textfont\bffam=\fourteenbf\scriptfont\bffam=\tenbf
\scriptscriptfont\bffam=\fivebf\def\bf{\fam\bffam\fourteenbf}%
\def\small{\twelvepoint}%
\def\large{\seventeenpoint}%
\def\normalbaselines{\lineskip1.44\normallineskip
  \baselineskip1.44\normalbaselineskip \lineskiplimit1.44\normallineskiplimit}
\setbox\strutbox=\hbox{\vrule height12.24pt depth5.04pt width0pt}%
\normalbaselines\rm}
\def\seventeenpoint{\def\rm{\fam0\seventeenrm}%
\textfont0=\seventeenrm\scriptfont0=\twelverm\scriptscriptfont0=\eightrm
\textfont1=\seventeeni\scriptfont1=\twelvei\scriptscriptfont1=\eighti
\textfont2=\seventeensy\scriptfont2=\twelvesy\scriptscriptfont2=\eightsy
\textfont3=\seventeenex\scriptfont3=\seventeenex
\scriptscriptfont3=\seventeenex
\textfont\itfam=\seventeenit\def\it{\fam\itfam\seventeenit}%
\textfont\slfam=\seventeensl\def\sl{\fam\slfam\seventeensl}%
\textfont\ttfam=\seventeentt\def\tt{\fam\ttfam\seventeentt}%
\textfont\bffam=\seventeenbf\scriptfont\bffam=\twelvebf
\scriptscriptfont\bffam=\eightbf
\def\bf{\fam\bffam\seventeenbf}%
\def\small{\fourteenpoint}%
\def\large{\seventeenpoint}%
\def\normalbaselines{\lineskip1.73\normallineskip
  \baselineskip1.73\normalbaselineskip \lineskiplimit1.73\normallineskiplimit}
\setbox\strutbox=\hbox{\vrule height14.7pt depth6.0pt width0pt}%
\normalbaselines\rm}
\def\eightpoint{\def\rm{\fam0\eightrm}%
\textfont0=\eightrm\scriptfont0=\sixrm\scriptscriptfont0=\fiverm
\textfont1=\eighti\scriptfont1=\sixi\scriptscriptfont1=\fivei
\textfont2=\eightsy\scriptfont2=\sixsy\scriptscriptfont2=\fivesy
\textfont3=\tenex\scriptfont3=\tenex\scriptscriptfont3=\tenex
\textfont\itfam=\eightit\def\it{\fam\itfam\eightit}%
\textfont\slfam=\eightsl\def\sl{\fam\slfam\eightsl}%
\textfont\ttfam=\eighttt\def\tt{\fam\ttfam\eighttt}%
\textfont\bffam=\eightbf\scriptfont\bffam=\sixbf
\scriptscriptfont\bffam=\fivebf\def\bf{\fam\bffam\eightbf}%
\def\small{\eightpoint}%
\def\large{\tenpoint}%
\def\normalbaselines{\lineskip.8\normallineskip
  \baselineskip.8\normalbaselineskip \lineskiplimit.8\normallineskiplimit}
\setbox\strutbox=\hbox{\vrule height7pt depth3pt width0pt}%
\normalbaselines\rm}
\def\small{\eightpoint}
\def\large{\twelvepoint}
\def\vfootnote#1{\insert\footins\bgroup
  \interlinepenalty\interfootnotelinepenalty
  \splittopskip\ht\strutbox 
  \splitmaxdepth\dp\strutbox \floatingpenalty\@MM
  \leftskip\z@skip \rightskip\z@skip \spaceskip\z@skip \xspaceskip\z@skip
  \footnotefont\textindent{#1}\footstrut\futurelet\next\fo@t}
\def\nfootnote{\advance\footnoteno1\@no=\footnoteno\advance\@no'140
\footnote{$^{\char\@no}$}}
\def\nvfootnote#1{\advance\footnoteno1\@no=\footnoteno\advance\@no'140
\def#1{$^{\char\@no}$}\vfootnote{$^{\char\@no}$}}
\def\titlefont{\seventeenpoint\rm}
\def\authorfont{\twelvepoint\rm}
\def\addressfont{\tenpoint\it}
\def\abstractfont{\eightpoint\rm}
\def\secfont{\tenpoint\bf}
\def\subsecfont{\tenpoint\sl}
\def\subsubsecfont{\tenpoint\it}
\def\eqnofont{\tenpoint\rm}
\def\footnotefont{\eightpoint\rm}
\catcode`@=12

\def\ve{\varepsilon}
\def\d{\partial}
\def\e{\mathinner{\rm e}}
\def\i{{\rm i}}

\def\Im{\mathop{\rm Im}}

\def\sh{\mathop{\rm sh}\nolimits}
\def\ch{\mathop{\rm ch}\nolimits}
\def\dn{\mathop{\rm dn}\nolimits}
\def\K{{\bf K}}

\def\half{{\textstyle{1\over2}}}
\def\H{{\cal H}}
\def\barbarF{\skew3\bar{\bar F}}
\def\barbarf{\skew6\bar{\bar f}}
\def\bc{\mathopen:}
\def\ec{\mathclose:}
\mathsurround=1pt


\reportno{LANDAU-97-TMP-2\\ \tt hep-th/9704148}
\Title{Scaling limit of the six vertex model in the framework\\
of free field representation}
\author{Michael Lashkevich}
\address{Landau Institute for Theoretical Physics, GSP-1,
117940 Moscow V-334, Russia}
\abstract{
The scaling limit of the spectrum, $S$ matrix, and of the form factors
of the polarization operator in the six vertex model has been found.
The result for the form factors is consistent
with the form factors of the sine-Gordon model found recently by Lukyanov.
We discuss the origin of the structure of the free field representation
for the sine-Gordon model at the critical coupling from the point of view of
the lattice model.
}
\date{April 1997}

\nref\ZZ{A.~B.~Zamolodchikov and Al.~B.~Zamolodchikov,
	{\it Annals in Phys.}\ {\bf 120}, 253 (1979)}
\nref\Smirnov{F.~A.~Smirnov, {\it Form Factors in Completely Integrable
	Models of Quantum Field Theory}, World Scientific, Singapore, 1992}
\nref\Lukbos{S.~Lukyanov,
	{\it Commun.\ Math.\ Phys.}\
	{\bf 167}, 183 (1995) {\tt hep-th/9307196}}
\nref\DFJMN{B.~Davies, O.~Foda, M.~Jimbo, T.~Miwa, and A.~Nakayashiki,
	{\it Commun.\ Math.\ Phys.}\ {\bf 151}, 89 (1993)}
\nref\JMMN{M.~Jimbo and K.~Miki, T.~Miwa, and A.~Nakayashiki,
	{\it Phys.\ Lett.}\ {\bf A168} 256 (1992)}
\nref\JM{M.~Jimbo, T.~Miwa, {\it Algebraic Analysis of Solvable Lattice
	Models}, CBMS Regional Conference Series in Mathematics,
	vol.\ 85, AMS, 1994}
\nref\Baxter{R.~J.~Baxter,
	{\it Exactly Solved Models in Statistical Mechanics},
	Academic Press, 1982}
\nref\Luther{A.~Luther, {\it Phys.\ Rev.}\ {\bf B14}, 2153 (1976)}
\nref\Lukexp{S.~Lukyanov, Form-factors of exponential fields
	in the sine-Gordon model, CLNS~97/1471, March 1997,
	{\tt hep-th/9703190}}
\nref\JKM{M.~Jimbo, H.~Konno, and T.~Miwa, Massless $XXZ$ model
	and degeneration of the elliptic algebra, October 1996,
	{\tt hep-th/9610079}}
\nref\LZ{S.~Lukyanov and A.~Zamolodchikov, Exact expectation values of
	local fields in quantum sine-Gordon model, CLNS~96/1444,
	\hbox{RU-96-107}, {\tt hep-th/9611238}}
\nref\LP{S.~Lukyanov and Ya.~Pugai, Multi-point local height probabilities
	in the integrable RSOS model, February 1996, {\tt hep-th/9602074}}
\nref\BK{R.~J.~Baxter and S.~B.~Kelland, {it J.\ Phys.}\ {\bf C7},
	L403 (1974)}
\nref\Bougourzi{A.~Abada, A.~H.~Bougourzi, and B.~Si-Lakhal,
	Exact four-spinon dynamical correlation function of the
	Heisenberg model, \hbox{ITP-SB-96-72}, December 1996,
	{\tt hep-th/9702028}}

\nsec Introduction

In the recent years a considerable progress has been made in investigation
of integrable models of two-dimensional quantum field theory.
$S$ matrices\refs{\ZZ} and form factors\refs{\Smirnov} have been found
in the framework of the bootstrap approach.
A free field representation\refs{\Lukbos}
for the form factors of the sine-Gordon and Thirring models was proposed.
But these achievements are based on some guesses and we need some additional
physical grounds to understand better the nature of these results.
The main problem can be formulated in the following way. It is possible
to construct sets of form factors, that describe some local operators
on the basis of the axiomatic approach. Moreover, it is possible construct
the whole space of local operators. But it turns out very difficult to
understand which operator corresponds to a given set of the form factors,
and vice versa: if we consider a given operator, then
which set of axiomatically
allowed form factors corresponds to it?

A approach to lattice models of statistical
mechanics based on the free field representation of correlation
functions and form factors
has been developed by the Kyoto group\refs{\DFJMN,\JMMN,\JM}.
In contrast to the situation in the quantum field theory
this approach has a firm physical basis.
There is a simple rule that allows one
to construct a bosonic representative to any
local lattice object. So it one can easily construct a set of form
factors corresponding to each local lattice insertion.
On the other hand,
the lattice models are known to be described by models of quantum field
theory at large scale in the vicinity of the critical point.
So, it seems promising to investigate these models
in the vicinity of the critical point to extract some information
on the respective quantum field models. Our main strategy is the following.
We consider a lattice theory and find form factors of some local insertions
in the scaling limit. Then from some additional information, such as
scaling dimensions, conservation laws, symmetries,
mutual locality {\it etc}. we
associate a local operator in the field theory to the insertion.
This allows us to find to relate form factors to particular operators.
Here we do not fulfil this program sequentially, but make some steps
in this direction.

In this letter we consider the six vertex model\refs{\Baxter}
in the antiferroelectric region in the vicinity of
the critical point. In the critical region this model
is described by the $SU(2)$ invariant Thirring model
or, equivalently, by the sine-Gordon model
at the critical coupling\refs{\Luther}.
We analyze the spectrum, the $S$ matrix,
and the form factors of the polarization operator $\sigma^z$. Our result
does not match that obtained by Jimbo and Miwa\refs{\JM}, but we shall see
that it is consistent with the formula for form factors in the sine-Gordon
model, guessed recently by Lukyanov\refs{\Lukexp}. We show the crucial role
of the zero mode operators to the behavior of the form factors in the
scaling limit, and show how to pass to a `conjugate' pair of the zero
mode operators simplifies the limit.

\nsec Spectrum and $S$ Matrix

The six vertex model in
the antiferroelectric region is described by two parameters $x$ and $\zeta$,
$$
0<x<1,\qquad 1<\zeta<x^{-2},
$$
so that the weight matrix of the model is given by the trigonometric
$R$ matrix
$$
\eqalign{
R(\zeta)_{++}^{++}
&=R(\zeta)_{--}^{--}=\kappa^{-1}(\zeta),
\\
R(\zeta)_{+-}^{+-}
&=R(\zeta)_{-+}^{-+}=\kappa^{-1}(\zeta){(\zeta-1)x\over1-x^2\zeta},
\\
R(\zeta)_{+-}^{-+}&
=R(\zeta)_{-+}^{+-}=\kappa^{-1}(\zeta){(1-x^2)\zeta^{1/2}\over1-x^2\zeta},
\qquad\smash{
\kappa(\zeta)
=\zeta^{1/2}{(x^4\zeta;x^4)_\infty(x^2\zeta^{-1};x^4)_\infty
\over(x^4\zeta^{-1};x^4)_\infty(x^2\zeta;x^4)_\infty},}
}\eqlabel\EQsvRmatrix
$$
where $(z;p)_\infty=\prod_{n=0}^\infty(1-zp^n)$. The $R$ matrix satisfies
the Yang--Baxter equation, and $\zeta$ serves as a multiplicative
spectral parameter. In particular, the row-to-row transfer matrices
$T(\zeta)\equiv T(x;\zeta)$ commute for different values of $\zeta$
but fixed $x$.

The eigenvectors of the transfer matrix at the infinite lattice
are `$N$-particle' states $|\varepsilon_1z_1,\ldots,\varepsilon_Nz_N\rangle$,
$$
T(\zeta)|\varepsilon_1z_1,\ldots,\varepsilon_Nz_N\rangle
=\prod_{j=1}^N\tau(z_j/\zeta)
|\varepsilon_1z_1,\ldots,\varepsilon_Nz_N\rangle
\eqlabel\EQeiginstates
$$
with $\varepsilon_j=\pm\equiv\pm1$ being the `charge' of `$j$th particle',
and $z_j$ is the `rapidity' of `$j$th particle',
$$
|z_j|=1.
\eqlabel\EQzvalues
$$
The `one-particle' eigenvalue function $\tau(z)$ is given by
$$
\tau(z)=z^{1/2}{\Theta_{x^4}(-x^3z)\over\Theta_{x^4}(-xz)},
\eqlabel\EQtaufunc
$$
where $\Theta_p(z)=(z;p)_\infty(p/z;p)_\infty(p;p)_\infty$ is a
multiplicative form of the $\theta_1$ function with the Jacobi parameter
$p^{1/2}$:
$$
\Theta_p(z)=-2\i p^{-1/8}z^{1/2}\theta_1(v;\tau),\quad z=\e^{2\pi\i v},
\quad \e^{\i\pi\tau}=p^{1/2}.
$$
Note that this form of the eigenvalue function is universal for
all lattice models based on the $A_1$ type $R$ matrices.

We know that in the scaling limit, when we are bringing the
system closer to the critical point, simultaneously enlarging the scale,
we have to obtain a system of relativistic particles
with the states
$|\ve_1\theta_1,\ldots,\ve_N\theta_N\rangle$,
the Hamiltonian $H$, and momentum operator $P$, such that
$$
\eqalign{
P\,|\ve_1\theta_1,\ldots,\ve_N\theta_N\rangle
&=M\sum_{j=1}^N\sh\theta_j\,|\ve_1\theta_1,\ldots,\ve_N\theta_N\rangle,
\\
H\,|\ve_1\theta_1,\ldots,\ve_N\theta_N\rangle
&=M\sum_{j=1}^N\ch\theta_j\,|\ve_1\theta_1,\ldots,\ve_N\theta_N\rangle,
}
\eqlabel\EQrelspectrum
$$
with $M$ being the mass of the particles, and $\theta_j$ being
rapidities. Here the parameter $\theta$
is connected somehow with $z$, and the functions
$M\ch\theta$ and $M\sh\theta$ must be related somehow with
the function $\tau(z)$. The scattering matrices of the lattice
models must as well become the relativistic scattering matrices
in the scaling limit.

The scaling limit corresponds to the limit $x\to1$, so we set
$$
x=e^{-\epsilon},\qquad\epsilon\to0.
$$
Let us introduce also additive spectral
parameters $\gamma$ and $\vartheta$
that survive in the scaling limit:
$$
\eqalign{
\span
\zeta=x^{-2\gamma/\pi},
\qquad
z=x^{-2\i\vartheta/\pi},
\\
\span
\gamma,\vartheta\in{\bf R},
\qquad
0\leq\gamma\leq\pi.
}\eqlabel\EQgammavarthetadef
$$

It is convenient to pass to the conjugate module in the theta functions.
Namely, for the function $\tau(\vartheta)$ we have
$$
\tau(\vartheta)
={\theta_4\left({\i\vartheta\over2\pi}+{1\over4};{\i\pi\over2\epsilon}\right)
\over
\theta_4\left({\i\vartheta\over2\pi}-{1\over4};{\i\pi\over2\epsilon}\right)}
=k'^{-1/2}\dn\left({\K\over2}-{\i\K\vartheta\over\pi}\right).
$$
The Jacobi parameter, corresponding to the theta functions is
$$
q=\e^{-\pi^2/2\epsilon}.
$$
We also use the usual designations: $k$ and $k'$ are module and conjugate
module, $\K$ and $\K'$ are half-periods.

The Jacobi parameter $q$ tends to zero in the scaling limit
and we can use the standard expansions in $q$ for theta functions.
But before
doing it let us understand, which values of the spectral
parameter correspond to the low-lying
excitations which survive at a large scale.
The low-lying excitation correspond to the largest in absolute value
transfer matrix eigenvalues. The physical values of the
spectral parameters $\zeta$ and $z$ satisfy the constraints
$1\leq\zeta\leq x^{-2}$ and $|z|=1$.
So we ought to
find the maximum of $|\tau(z/\zeta)|^2$ in $z$ on the unit circle
for fixed $\zeta$ in the interval $1\leq\zeta\leq x^{-2}$, or,
in other words, the maximum of $|\tau(\vartheta+\i\gamma)|^2$
in real $\vartheta$ for fixed $\gamma$, $0\leq\gamma\leq\pi$.
To do it, note that the function
$$
\mathopen|\tau(\vartheta+\i\gamma)\mathclose|^2
={\theta_4\left({\i\vartheta\over2\pi}-{\gamma\over2\pi}+{1\over4};
{\i\pi\over2\epsilon}\right)
\theta_4\left({\i\vartheta\over2\pi}+{\gamma\over2\pi}-{1\over4};
{\i\pi\over2\epsilon}\right)
\over
\theta_4\left({\i\vartheta\over2\pi}-{\gamma\over2\pi}-{1\over4};
{\i\pi\over2\epsilon}\right)
\theta_4\left({\i\vartheta\over2\pi}+{\gamma\over2\pi}+{1\over4};
{\i\pi\over2\epsilon}\right)}
\eqlabel\EQtauabssqr
$$
of $\vartheta$
can be analytically continued to a doubly periodic
function on the complex plane. Then its
derivative in $\vartheta$ has two double poles in the
periodicity region, and, therefore,
has exactly 4 zeros. Using the symmetries of this function it
is easy to check that all 4 extremal values are situated at
the `symmetric points' and two of them on the real axis.
The maximum is achieved at the point $\vartheta=0$ and the minimum at
$\vartheta=\pi^2/2\epsilon$.

We see that we ought to consider the limit of the elliptic functions
$q\to0$ ($\K\to\pi/2$, $\K'\to\infty$) for the values of $\vartheta$
in any finite interval around $0$.
Applying to (\EQtauabssqr) the standard limiting formulas
$$
\eqalign{
\span
\dn{2\K\alpha\over\pi}={\pi\over2\K}[1+4q\cos2\alpha+O(q^2)],
\\
\span
k'=1-8q+O(q^2),\quad\K={\pi\over2}[1+4q+O(q^2)],
}
$$
we obtain
$$
\tau(\vartheta)=1+4\i q\sh\vartheta+O(q^2).
$$
The system can be considered continuous if
$$
q\e^{\mathopen|\vartheta\mathclose|}\ll1
\quad{\rm or}\quad
-{\pi^2\over2(1+\delta)\epsilon}<\vartheta<{\pi^2\over2(1+\delta)\epsilon},
\quad\delta>0.
$$
Let us think that our system lives on the lattice $(am,bn)$, $m,n\in{\bf Z}$,
$a$ and $b$ being lattice parameters.
We will think of $am$ as of the imaginary time $\i t'$ and of
$bn$ as of the space coordinate $x'$. Recall that the operator
$T(1)$ is the shift operator by $b$\refs{\JM}.
So the momentum operator $p'$ is given by
$$
\e^{\i p'(\vartheta)b}\simeq1+4\i q\sh\vartheta
$$
or
$$
p'(\vartheta)={4q\over b}\sh\vartheta.
$$
The Hamiltonian is given by $\e^{-aH}=T(z)$, and for the one-particle
energy we have
$$
\e^{-E'(\vartheta)a}=\tau(\vartheta+\i\gamma)
\simeq1+4\i q\sh(\vartheta+\i\gamma)
$$
or
$$
E'(\vartheta)=-{4\i q\over a}\sh(\vartheta+\i\gamma)
={4q\over a}(\sin\gamma\ch\vartheta-\i\cos\gamma\sh\vartheta).
$$
To restore the relativistic spectrum we recall that the initial
lattice system is essentially anisotropic. Hence, we need to
change coordinates $(t',x')\to(t,x)$, so that the spectrum would
take the standard form
$E(\theta)=M\ch\theta$, $p(\theta)=M\sh\theta$.
It is achieved by the coordinate transformation
$$
\left\{
\matrix{
t=t'{b\over a}\sin\gamma\hfill
&=-\i b\sin\gamma\cdot m,\hfill
\\
x=\i t'{b\over a}\cos\gamma+x'\hfill
&=+b\cos\gamma\cdot m+bn,\hfill
}\right.
\eqlabel\EQcoordchange
$$
and by setting
$$
\theta=\vartheta,\qquad M={4q\over b}.
\eqlabel\EQthetaM
$$

Consider now the $S$ matrix. For the six vertex model
it is given by\refs{\JM}
$$
S(z)=-R(z^{-1}).
$$
In the scaling limit we obtain
$$
\eqalign{
S(\theta)_{++}^{++}
&=S(\theta)_{--}^{--}=-S_0(\theta),
\\
S(\theta)_{+-}^{+-}
&=S(\theta)_{-+}^{-+}=-S_0(\theta){\theta\over\i\pi-\theta},
\\
S(\theta)_{+-}^{-+}
&=S(\theta)_{-+}^{+-}=-S_0(\theta){\i\pi\over\i\pi-\theta},
\qquad\smash{
S_0(\theta)
={\Gamma\left(1+{\i\theta\over2\pi}\right)
\Gamma\left({1\over2}-{\i\theta\over2\pi}\right)
\over\Gamma\left(1-{\i\theta\over2\pi}\right)
\Gamma\left({1\over2}+{\i\theta\over2\pi}\right)}.}
}
$$
This $S$ matrix coincides with the critical sine-Gordon $S$ matrix.%
\nfootnote{In the case of the six vertex model we do not need the
duality transformation here\refs{\JKM}. }

The sine-Gordon model is the model with the action
$$
{\cal A}=\int d^2x\,\left({1\over16\pi}(\d_\mu\varphi)^2
+{\lambda\over\beta^2}\cos\beta\varphi\right).
$$
The dimensionality of the constant $\lambda$ is $M^{2-2\beta^2}$.
For $0<\beta<1$ the term $\cos\beta\varphi$
is a relevant perturbation, and
the theory is superrenormalizable. For $\beta>1$ the perturbation is
irrelevant, and the properties of the theory are completely unknown.
The critical sine-Gordon theory corresponds to the marginal value $\beta=1$.
In fact, dimensional transmutation takes place at this point,
and the theory is an integrable theory
of massive interactive charged solitons. $S(\theta)$ is the pair
soliton--soliton $S$ matrix.

\nsec Free Field Representation and Conjugate Zero Mode

Now let us turn to the calculation of form factors. We start from
the free field representation of Ref.~\Refs{\JM}.

Consider the Heisenberg algebra
$$
[P,Q]=-\i,
\qquad
[a_m,a_n]=\delta_{m+n,0}{[2m]_x[m]_x\over m}
\qquad
\left([s]_x={x^s-x^{-s}\over x-x^{-1}}\right),
$$
The spaces $\H_i$, $i=0,1$ are generated
by the operators $a_{-n}$, $n>0$, and $\exp(\pm\i\sqrt2 Q)$
from the vacuum vectors
$|0\rangle^{(i)}$, defined as
$$
\eqalign{
a_n|0\rangle^{(i)}&=0,\quad n>0,
\\
P|0\rangle^{(i)}&={i\over\sqrt2}|0\rangle^{(i)}.
}
$$
We also shall use the notation $|p\rangle$ for the vector
$$
|p\rangle=\e^{\i pQ}|0\rangle^{(0)},
\qquad
P\,|p\rangle=p\,|p\rangle.
$$

The vertex operators $V_\ve^{(1-i,i)}(\zeta)$, $\tilde V_\ve^{(1-i,i)}(z)$:
$\H_i\rightarrow\H_{1-i}$ are defined as\refs{\JM}%
\nfootnote{In terms of Ref.\ \Refs{\JM}
$V_\ve^{(1-i,i)}(\zeta)\sim\Phi_{-\ve}^{(1-i,i)}(\zeta^{1/2})$,
$\tilde V_\ve^{(1-i,i)}(z)\sim\Psi_{-\ve}^{*(1-i,i)}(z)$.}
$$
\eqalign{
V_+^{(1-i,i)}(\zeta)
&=G^{1/2}x^{3/4} \bc\e^{\i\phi(\zeta)}\ec,
\\
V_-^{(1-i,i)}(\zeta)
&=G^{1/2}x^{3/4}\oint_{C_1}{dv\over2\pi\i}\,
F_1(v,\zeta)
\bc\e^{\i\phi(\zeta)-\i\phi(x^{-1}v)-\i\phi(xv)}\ec,
\\
\tilde V_+^{(1-i,i)}(z)
&=G^{-1/2}x^{3/4}z^{1/2}\bc\e^{\i\tilde\phi(z)}\ec,
\\
\tilde V_-^{(1-i,i)}(z)
&=G^{-1/2}x^{3/4}z^{1/2}\oint_{C_2}{dw\over2\pi\i}\,
F_2(w,z)
\bc\e^{\i\tilde\phi(z)-\i\tilde\phi(x^{-1}w)-\i\tilde\phi(xw)}\ec,
}\eqlabel\EQvvdef
$$
where
$$
G={(x^2;x^4)_\infty\over(x^4;x^4)_\infty},
\qquad
F_1(v,\zeta)={(1-x^2)v\zeta^{1/2}\over x(v-x^{-1}\zeta)(v-x\zeta)},
\qquad
F_2(w,z)={(1-x^2)z^{1/2}\over x(w-x^{-1}z)(w-xz)},
\eqlabel\EQgFFdef
$$
and
$$
\eqalign{
\phi(\zeta)
&={1\over\sqrt2}(Q-\i P\,\log(x^3\zeta))
+\sum_{n\neq0}{a_n\over\i[2n]_x}x^{|n|/2}\zeta^{-n},
\\
\tilde\phi(z)
&=-{1\over\sqrt2}(Q-\i P\,\log(x^3z))
-\sum_{n\neq0}(-1)^{n}{a_n\over\i[2n]_x}x^{-|n|/2}z^{-n}.
}\eqlabel\EQphidef
$$
The normal ordering operation $\bc\ldots\ec$ places $P$ to the right
of $Q$, and $a_n$ to the right of $a_{-n}$, $n>0$. The contours $C_1$ and
$C_2$ encircle the point $0$ so that $x\zeta$ is inside and $x^{-1}\zeta$ is
outside $C_1$, and $x^{-1}z$ is inside and $xz$ is outside $C_2$.

The form factors of local operators $O$ are given by\refs{\JM}
$$
\langle0|O|\ve_1\zeta_1\ldots\ve_N\zeta_N\rangle^{(i)}_{\rm in}
={1\over\chi_i(x^4)}
\Tr_{\H_i}\left(x^{4D^{(i)}}\hat O^{(i,i')}
\tilde V_{\ve_N}^{(i',1-i')}(z_N)\ldots\tilde V_{\ve_1}^{(1-i,i)}(z_1)\right),
\eqlabel\EQtrformula
$$
Here
$$
D^{(i)}={P^2\over2}-{P\over2\sqrt2}
+\sum_{n=1}^\infty{n^2\over[2n]_x[n]_x}a_{-n}a_n
\eqlabel\EQgrading
$$
is the shift operator
$$
[D^{(i)},V_\ve^{(1-i,i)}(\zeta)]=\zeta{d\over d\zeta}V_\ve^{(1-i,i)}(\zeta),
\qquad
[D^{(i)},\tilde V_\ve^{(1-i,i)}(z)]=z{d\over dz}\tilde V_\ve^{(1-i,i)}(z),
$$
and $\chi_i(t)$ is the character of $\H_i$,
$$
\chi_i(t)=\Tr_{\H_i}t^{D^{(i)}}=(t^{1/2};t)_\infty^{-1}.
$$
The operator $\hat O^{(i,i')}$ is determined by the operator $O$ and
consists of the operators $V_\ve^{(1-i,i)}(\zeta)$. We shall consider
here an important example of the polarization operator
$\sigma^z=\pmatrix{1&0\\0&-1}$,
inserted into one of the vertical links of the lattice. The respective
operator $\widehat{(\sigma^z)}^{(i,i)}$ is given by
$$
\widehat{(\sigma^z)}^{(i,i)}
=-\sum_\ve \ve V_{-\ve}^{(i,1-i)}(x^{-2}\zeta)V_\ve^{(1-i,i)}(\zeta).
$$

Consider an arbitrary trace
$$
T^{(i)}_N={1\over\chi_i(x^4)}
\Tr_{\H_i}\left(x^{4D^{(i)}}\bc\e^{\phi_N}\ec\ldots\bc\e^{\phi_1}\ec\right),
$$
where
$$
\bc\e^{\phi_j}\ec=\e^{\alpha_jQ}\e^{\epsilon\beta_jP}
\exp\biggl(\sum_{n=1}^\infty A_na_{-n}\biggr)
\exp\biggl(\sum_{n=1}^\infty A_{-n}a_n\biggr).
$$
It is easy to check that the trace can be factorized,
$$
T^{(i)}_N=H^{(i)}_N F_N,
$$
into the contribution of the zero mode $(Q,P)$, $H^{(i)}_N$,
and the contribution of the oscillators $a_n$, $F_N$. The oscillators
contribution is given by
$$
\eqalign{
F_N&=\prod_j C_j \prod_{j<j'}F_{jj'},
\\
\log C_j&={1\over\chi_*(x^4)}\Tr_*\left(x^{4D_*}\phi_{+,j}\phi_{-,j}\right),
\\
\log F_{jj'}&={1\over\chi_*(x^4)}
\Tr_*\left(x^{4D_*}\phi_{*j}\phi_{*j'}\right),
\\
\chi_*(t)&=\Tr_*\left(t^{D_*}\right).
}\eqlabel\EQnzfunc
$$
Here asterisks mean that the contribution of oscillators is only taken into
account; $\phi_+$ and $\phi_-$ are contributions into $\phi$
from $a_{-n}$ and $a_n$ respectively ($n>0$).

The zero mode contribution is the most interesting part.
Let us
write it down explicitly.
It is only nonzero for $\sum_j\alpha_j=0$. Let us introduce the notations
$$
A=\sum_{j<j'}^N\alpha_j\beta_{j'},
\qquad
B=\sum_{j=1}^N\beta_j.
\eqlabel\EQABdef
$$
Then
$$
\eqalignno{
H^{(i)}_N\chi_0(x^4)
&=\Tr_0^{(i)}\left(x^{4D_0^{(i)}}\e^{\alpha_NQ}\e^{\epsilon\beta_NP}
\ldots\e^{\alpha_1Q}\e^{\epsilon\beta_1P}\right)
\\
&=\sum_{n\in{\bf Z}+i/2}
\left\langle n\sqrt2\right|\,
\e^{-4\epsilon\left({P^2\over2}-{P\over2\sqrt2}\right)}
\e^{\alpha_NQ}\e^{\epsilon\beta_NP}\ldots\e^{\alpha_1Q}\e^{\epsilon\beta_1P}
\,\left|n\sqrt2\right\rangle
\\
&=\e^{-\i\epsilon A}
\sum_{n\in{\bf Z}+i/2}\exp\biggl(-4\epsilon\left(n^2-{n\over2}\right)
+\epsilon n\sqrt2\sum_j\beta_j\biggr)
=\e^{-\i\epsilon A}
\theta_{3-i}\biggl(-\i\epsilon{2+\sqrt2 B\over2\pi};\i{4\epsilon\over\pi}
\biggr)
\\
&={1\over2}\sqrt{\pi\over\epsilon}
\e^{-\i\epsilon A}
\exp\left[4\epsilon\left({1\over4}+{B\over4\sqrt2}\right)^2\right]
\theta_{3+i}\left({1\over4}+{B\over4\sqrt2};\i{\pi\over4\epsilon}
\right),
\lnlabel\EQzfunc}
$$
where the subscript $0$ means the contribution of the zero mode.
We passed to a conjugate module in the last equality. The most interesting
feature of this expression is that we can rewrite the last theta function
as a trace of an operator. Expanding it in the series, we can easily check
that
$$
\eqalignno{
H^{(i)}_N\chi_0(x^4)
&={1\over2}\sqrt{\pi\over\epsilon}
\exp\left[4\epsilon\left({1\over4}+{B\over4\sqrt2}\right)^2\right]
\\
&\quad\times
\sum_{n\in{\bf Z}}
\,\,\bigg.'\!\!\left\langle{\pi n\over2\sqrt2}\right|
\e^{-{4\over\epsilon}{P^{\prime2}\over2}+\i(-1)^i\sqrt2 P'}
\e^{-\i\epsilon\alpha_N Q'}\e^{\i\beta_N P'}\ldots
\e^{-\i\epsilon\alpha_1 Q'}\e^{\i\beta_1 P'}
\left|{\pi n\over2\sqrt2}\right\rangle'.
\lnlabel\EQcpq}
$$
Here $P'$ and $Q'$ is a new pair of canonically conjugate operators
and $|p\rangle'$ designates the eigenvector of $P'$:
$$
[P',Q']=-\i,
\qquad
P'|p\rangle'=p\,|p\rangle',
\qquad
|p\rangle'=\e^{\i pQ}|0\rangle'.
\eqlabel\EQcpqcommut
$$
Naively, the term with $n=0$ only survives in the limit $\epsilon\to0$.
But as we shall see below in the calculation of the form factors of
$\sigma^z$ this term vanishes, so that
the terms with $n=\pm1$ become leading.

Note also that we can substitute $P\to\i P'/\epsilon$ and
$Q\to-\i\epsilon Q'$ directly in the vertex operators. This substitution
does not affect the commutation relations of the vertex operators,
but the summation rule over the momentum values changes drastically
and the operator $D^{(i)}$ must be substituted by
$$
D^{\prime(i)}=\epsilon^{-2}\left({P'^2\over2}-\i{(-)^i P'\over2\sqrt2}\right)
+\sum_{n=1}^\infty{n^2\over[2n]_x[n]_x}a_{-n}a_n.
\eqlabel\EQDprimedef
$$

The explicit form of the nonzero `two-point' traces is
$$
\eqalign{
-\log F_{11}(\gamma_1-\gamma_2)
&={1\over\chi_*(x^4)}
\Tr_*\left(x^{4D_*}\phi_*(\zeta_1)\phi_*(\zeta_2)\right)
=\sum_{n=1}^\infty
{\e^{-\epsilon n}\over n}{\sh\epsilon n\over\sh^2 2\epsilon n}
\ch 2\epsilon n\left(1-{\gamma_1-\gamma_2\over\pi}\right),
\\
-\log F_{22}(\theta_1-\theta_2)
&={1\over\chi_*(x^4)}
\Tr_*\left(x^{4D_*}\tilde\phi_*(z_1)\tilde\phi_*(z_2)\right)
=\sum_{n=1}^\infty
{\e^{\epsilon n}\over n}{\sh\epsilon n\over\sh^2 2\epsilon n}
\ch 2\epsilon n\left(1-\i{\theta_1-\theta_2\over\pi}\right),
\\
-\log F_{12}(\gamma_1-\i\theta_2)
&={1\over\chi_*(x^4)}
\Tr_*\left(x^{4D_*}\phi_*(\zeta_1)\tilde\phi_*(z_2)\right)
=\sum_{n=1}^\infty
{(-1)^{n-1}\over n}{\sh\epsilon n\over\sh^2 2\epsilon n}
\ch 2\epsilon n\left(1-{\gamma_1-\i\theta_2\over\pi}\right).
}\eqlabel\EQtptr
$$
Similarly, the traces of the form
$\chi_*^{-1}(x^4)\Tr_*(x^{4D_*}\phi_+\phi_-)$ are
$$
\eqalign{
-\log[x^{-3/4}G^{-1/2}C_+]
&=\sum_{n=1}^\infty{\e^{-3\epsilon n}\sh\epsilon n
\over2n\sh^2 2\epsilon n},
\\
-\log[x^{-3/4}G^{-1/2}F_1^{-1}(v,\zeta)C_-(\gamma,\mu)]
&=\sum_{n=1}^\infty{\e^{-3\epsilon n}\sh\epsilon n
\over2n\sh^2 2\epsilon n}
\left[-4\left(\ch2\epsilon n{\mu\over\pi}-1\right)\ch\epsilon n
+(2\ch\epsilon n-1)^2\right],
\\
-\log[x^{-3/4}G^{1/2}\tilde C_+]
&=\sum_{n=1}^\infty{\e^{-\epsilon n}\sh\epsilon n
\over2n\sh^2 2\epsilon n},
\\
-\log[x^{-3/4}G^{1/2}F_2^{-1}(w,z)\tilde C_-(\theta,\nu)]
&=\sum_{n=1}^\infty{\e^{-\epsilon n}\sh\epsilon n
\over2n\sh^2 2\epsilon n}
\left[-4\left(\ch2\epsilon n{\i\nu\over\pi}-1\right)\ch\epsilon n
+(2\ch\epsilon n-1)^2\right],
\\
\span
v=\e^{{2\epsilon\over\pi}(\gamma+\mu)},
\qquad
w=\e^{{2\i\epsilon\over\pi}(\theta+\nu)}.
}\eqlabel\EQctr
$$
It is convenient also to introduce the functions
$$
\eqalign{
\bar F_{11}(\gamma)&=F_{11}^{-1}(\gamma-\pi/2)F_{11}^{-1}(\gamma+\pi/2),
\\
\bar F_{12}(\gamma)&=F_{12}^{-1}(\gamma-\pi/2)F_{12}^{-1}(\gamma+\pi/2),
\\
\bar F_{22}(\theta)&=F_{22}^{-1}(\theta+\i\pi/2)F_{22}^{-1}(\theta-\i\pi/2),
}
\qquad
\eqalign{
\barbarF_{11}(\gamma)
&=\bar F_{11}^{-1}(\gamma-\pi/2)\bar F_{11}^{-1}(\gamma+\pi/2),
\\
\barbarF_{12}(\gamma)
&=\bar F_{12}^{-1}(\gamma-\pi/2)\bar F_{12}^{-1}(\gamma+\pi/2),
\\
\barbarF_{22}(\theta)
&=\bar F_{22}^{-1}(\theta+\i\pi/2)\bar F_{22}^{-1}(\theta-\i\pi/2),
}\eqlabel\EQfbar
$$
All traces of the vertex operators can be written in terms of the functions
(\EQtptr--\EQfbar). To do it we need to write a function $C_\ve$,
$z_j^{1/2}\tilde C_\ve$ for each $V_\ve(\zeta_j)$, $\tilde V_\ve(z_j)$,
a function
$F_{11}(\gamma_j-\gamma_{j'})$ for each pair $V_+(\zeta_j)$ and
$V_+(\zeta_{j'})$, $j>j'$, a function
$F_{11}(\gamma_j-\gamma_{j'})\bar F_{11}(\gamma_j-\gamma_{j'}-\mu_{j'})$
for each pair $V_+(\zeta_j)$ and $V_-(\zeta_{j'})$, $j>j'$, {\it etc.},
then to write the overall coefficient from the zero mode, and finally to
integrate over all variables $\mu_j$ and $\nu_j$.

\nsec Scaling Limit of Form Factors

Consider the scaling limit of the functions (\EQtptr). Applying the formula
$$
\sum_{n=1}^\infty\left({a_2\over\epsilon n^2}+{a_1\e^{-\epsilon n}\over n}
+\epsilon f(\epsilon n)\right)
={\pi^2\over6\epsilon}a_2-a_1\log\epsilon
+\int_0^\infty d\eta\,f(\eta)+O(\epsilon)
$$
for $f(\eta)$ being finite as $\eta\to0$, we easily obtain
$$
\eqadvance\EQFtof
\eqalignno{
F_{11}(\gamma)
&=\epsilon^{-1/4} q^{1/12}(f_{11}(\gamma)+O(\epsilon)),
&(\EQFtof\rm a)
\\
F_{22}(\theta)
&=\epsilon^{1/4} q^{1/12}(f_{22}(\theta)+O(\epsilon)),
&(\EQFtof\rm b)
\\
F_{12}(\gamma)
&=q^{1/24}\left(\exp{\epsilon\over4}\left[\left(1-{\gamma\over\pi}\right)^2
-{7\over12}\right]+O(\epsilon^\infty)\right),
&(\EQFtof\rm c)}
$$
with
$$
\eqalign{
-\log f_{11}(\gamma)
&=\int_0^\infty{d\eta\over\eta}\,
\left[\e^{-\eta}{\sh\eta\over\sh^2 2\eta}
\ch2\eta\left(1-{\gamma\over\pi}\right)
-{1\over4\eta}+{1\over4}\e^{-\eta}\right],
\\
-\log f_{22}(\theta)
&=\int_0^\infty{d\eta\over\eta}\,
\left[\e^{\eta}{\sh\eta\over\sh^2 2\eta}
\ch2\eta\left(1-\i{\theta\over\pi}\right)
-{1\over4\eta}-{1\over4}\e^{-\eta}\right].
}\eqlabel\EQfc
$$
The expression for $F_{12}$ demands a more detailed consideration.
It is easy to check, that the formula is correct up to all orders in
$\epsilon$, because in the decomposition of the summand
in $\epsilon$ the odd powers of $n$ vanish. And it
is easy to check that $\sum_{n=1}^\infty(-1)^{n-1} n^{2p}=0$, $p\geq1$.
However, we need a more strong (and more rigorous)
estimation than $O(\epsilon^\infty)$. We can give it for the ratio
$$
{\bar F_{12}(\gamma+\pi/2)\over\bar F_{12}(\gamma-\pi/2)}
={\Theta_{x^4}(-xy)\over\Theta_{x^4}(-x^3y)}
=y^{1/2}[1+O(q)],
\qquad
y=x^{2(1-\gamma/\pi)}.
\eqlabel\EQratio
$$
The last equality follows from the earlier consideration of the
function $\tau(z)$. We shall see that this estimation is strong enough to
find the form factors of $\sigma^z$.

For $C_\pm$, $\tilde C_\pm$ we find
$$
\eqalign{
C_+&\simeq\epsilon^{-1/8}q^{1/24}c_+,
\\
v C_-(\gamma,\mu)&
\simeq\epsilon^{-1/8}q^{1/24}c_-
{\pi\over2\epsilon}f_1(\mu)\e^{\epsilon\gamma/\pi},
\\
\tilde C_+&\simeq\epsilon^{-3/8}q^{1/24}\tilde c_+,
\\
w \tilde C_-(\theta,\nu)
&\simeq\epsilon^{-3/8}q^{1/24}\tilde c_-
{\pi\over2\epsilon}f_2(\nu)
}\eqlabel\EQCtoc
$$
with
$$
\eqalign{
\log c_+
&=-{1\over2}\int_0^\infty{d\eta\over\eta}
\left(\e^{-3\eta}{\sh\eta\over\sh^2 2\eta}
-{1\over4\eta}+{1\over4}\e^{-\eta}+{\e^{-\eta}\over2\ch\eta}\right),
\\
\log c_-
&=-{1\over2}\int_0^\infty{d\eta\over\eta}
\left(\e^{-3\eta}{\sh\eta\over\sh^2 2\eta}(2\ch\eta-1)^2
-{1\over4\eta}+{1\over4}\e^{-\eta}+{\e^{-\eta}\over2\ch\eta}\right),
\\
\log\tilde c_+
&=-{1\over2}\int_0^\infty{d\eta\over\eta}
\left(\e^{-\eta}{\sh\eta\over\sh^2 2\eta}
-{1\over4\eta}+{1\over4}\e^{-\eta}-{\e^{-\eta}\over2\ch\eta}\right),
\\
\log\tilde c_-
&=-{1\over2}\int_0^\infty{d\eta\over\eta}
\left(\e^{-\eta}{\sh\eta\over\sh^2 2\eta}(2\ch\eta-1)^2
-{1\over4\eta}+{1\over4}\e^{-\eta}-{\e^{-\eta}\over2\ch\eta}\right),
\\
f_1(\mu)
&={\pi\over(\mu-\pi/2)(\mu+\pi/2)}
\exp\int_0^\infty{d\eta\over\eta}\,{\e^{-3\eta}\over\sh2\eta}
\left(\ch2\eta{\mu\over\pi}-1\right),
\\
f_2(\nu)
&=-{\pi\over(\nu+\i\pi/2)(\nu-\i\pi/2)}
\exp\int_0^\infty{d\eta\over\eta}\,{\e^{-\eta}\over\sh2\eta}
\left(\ch2\eta{\i\nu\over\pi}-1\right).
}\eqlabel\EQcf
$$
All equalities in (\EQCtoc) are up to $O(\epsilon)$,
but the expression for $C_-$ includes the precise dependence on $\gamma$,
which will be important later.

We need to make one more remark. The exact formulas for form factors
on the lattice contain integrations over closed paths encircling zero.
It means that we must integrate over $\mu$ and $\nu$
along to infinite contours. But at $|\mu|,|\nu|\sim1/\epsilon$ the
formulas (\EQFtof--\EQcf) is no more valid. We should analyze
its contribution.
We can argue in the following way. It is easy to
check that
$$
\eqalign{
f_{11}(\i\theta)
&\sim|\theta|^{-1/4}\e^{|\theta|/4},
\\
f_{22}(\theta)
&\sim|\theta|^{1/4}\e^{|\theta|/4},
}
\qquad
\eqalign{
f_1(\i\theta)
&\sim|\theta|^{-1/2}\e^{-|\theta|/2},
\\
f_2(\theta)
&\sim|\theta|^{-3/2}\e^{-|\theta|/2}
}
$$
for large real $\theta$.
Therefore for large imaginary $\mu$ or large real $\nu$ a factor
$\sim f_1(\mu)f_{11}^{-2}(\mu)\sim\e^{-|\mu|}$
or $\sim f_2(\nu)f_{22}^{-2}(\nu)\sim|\nu|^{-2}\e^{-|\nu|}$ appears.
We shall see that the zero mode gives the factors $\sim\e^{|\mu|}$,
$\e^{|\nu|}$ in the case of $\sigma^z$. So the contribution of
$\nu_j\sim1/\epsilon$ is negligible, but the contribution of
$\mu\sim1/\epsilon$ must be taken into account. Nevertheless,
as we shall discuss later, the correct answer can be extracted from
the approximation for small $\mu$.

At last, we define the states
$$
|\ve_1\theta_1,\ldots,\ve_N\theta_N\rangle^{(i)}_{\rm in}
=\left(2\epsilon\over\pi\right)^{N/2}
|\ve_1 z_1,\ldots,\ve_N z_N\rangle^{(i)}_{\rm in}
$$
so that it would have the natural for
two-dimensional field theory normalization:
$$
{1\over N!}\int{d^N\theta\over(2\pi)^N}
\,{}^{(i)}_{\rm in}\!\langle\ve_1\theta'_1,\ldots,\ve_N\theta'_N
|\ve_1\theta_1,\ldots,\ve_N\theta_N\rangle^{(i)}_{\rm in}
=1.
$$

Now we are ready to calculate of the form factors of $\sigma^z$
in the scaling limit. The form factors of $\sigma^z$ read precisely
$$
\langle0|\sigma^z
|\ve_1\theta_1,\ldots,\ve_{2n}\theta_{2n}\rangle_{\rm in}^{(i)}
=-\int_{C'_1}{d\mu\over2\pi\i}\int_{C'_2}{d^n\nu\over(2\pi)^n}
\sum_{\alpha=\pm}\alpha H_{-\alpha\alpha}^{(i)}(\gamma,\mu;\ve,\theta,\nu)
F_{-\alpha\alpha}(\gamma,\mu;\ve,\theta,\nu),
\eqlabel\EQintegral
$$
where $H$ and $F$ are contributions of the zero mode and the oscillators
respectively.
Let us begin with the oscillators.

The contour $C'_1$ goes from $-\i\infty$ to $+\i\infty$ to the right
of $-\half\pi$ and to the left of $\half\pi$,
and the contour $C'_2$ goes from $-\infty$ to
$+\infty$ above ${\i\over2}\pi$ and below $-{\i\over2}\pi$
(with an inflection).

The oscillators contribution factorizes as follows:
$$
F_{-\alpha\alpha}(\gamma,\mu;\ve,\theta,\nu)
={2\epsilon\over\pi}F_{-\alpha\alpha}(\mu) F(\ve,\theta,\nu)
S_{-\alpha\alpha}(\gamma,\mu;\ve,\theta,\nu).
$$
Here $F_{-\alpha\alpha}(\mu)$ is the $F$ function for
the vacuum expectation of $\sigma^z$:
$$
\eqalign{
F_{-+}(\mu)&={2\epsilon\over\pi}C_-(\gamma+\pi,\mu)C_+
F_{11}(\pi)\bar F_{11}(\mu+\pi),
\\
F_{+-}(\mu)&={2\epsilon\over\pi}C_+C_-(\gamma,\mu)
F_{11}(\pi)\bar F_{11}(\pi-\mu).
}\eqlabel\EQFone
$$
>From the evident property $F_{11}(\gamma+2\pi)=F_{11}(-\gamma)$ we obtain
$$
{F_{+-}(\mu)\over F_{-+}(\mu)}
={C_-(\gamma)\over C_-(\gamma+\pi)}=x=\e^{-\epsilon}.
\eqlabel\EQFratio
$$
Note that this result is precise.
Up to $O(\epsilon)$ the quantities $F_{-+}(\mu)$ and $F_{+-}(\mu)$
coincide and
$$
F_{-+}(\mu)\simeq c_+c_-f_1(\mu)f_{11}(\pi)\bar f_{11}(\mu+\pi).
\eqlabel\EQapprox
$$

The function $F(\ve,\theta,\nu)$ is the factor that only depends on
the asymptotic states rather than on the local operator:
$$
\eqalignno{
F(\ve,\theta,\nu)
&=\left(2\epsilon\over\pi\right)^{2n}\e^{\i{\epsilon\over\pi}\sum\theta_j}
\tilde C_+^n\prod_{j\in N_-}\tilde C_-(\theta_j,\nu_j)
\prod_{j<j'}F_{22}(\theta_{j'}-\theta_j)
\\
&\quad\times
\prod_{j<j',j'\in N_-}\bar F_{22}(\theta_{j'}-\theta_j+\nu_{j'})
\prod_{j<j',j\in N_-}\bar F_{22}(\theta_{j'}-\theta_j-\nu_j)
\prod_{j<j'\in N_-}\barbarF_{22}(\theta_{j'}-\theta_j+\nu_{j'}-\nu_j)
\\
&\simeq
\left(2\over\pi\right)^n\tilde c_+^n\tilde c_-^n
\prod_{j\in N_-} f_2(\theta_j,\nu_j)
\prod_{j<j'}f_{22}(\theta_{j'}-\theta_j)
\\
&\quad\times
\prod_{j<j',j'\in N_-}\bar f_{22}(\theta_{j'}-\theta_j+\nu_{j'})
\prod_{j<j',j\in N_-}\bar f_{22}(\theta_{j'}-\theta_j-\nu_j)
\\
&\quad\times
\prod_{j<j'\in N_-}\barbarf_{22}(\theta_{j'}-\theta_j+\nu_{j'}-\nu_j)
\lnlabel\EQFtwo}
$$
with $N_\pm=\{j|\ve_j=\pm\}$. From the requirement $\sum_j\alpha_j=0$
we have $\sum_j\ve_j=0$ or $\#N_+=\#N_-$.

The third factor $S_{-\alpha\alpha}(\gamma,\mu;\ve,\theta,\nu)$ is given by
$$
\eqalign{
S_{-+}(\gamma,\mu;\ve,\theta,\nu)
&=\prod_{j}F_{12}(\gamma-\i\theta_j+\pi)F_{12}(\gamma-\i\theta)
\bar F_{12}(\gamma-\i\theta_j+\mu+\pi)
\\
&\quad\times
\prod_{j\in N_-}\bar F_{12}(\gamma-\i\theta_j-\i\nu_j+\pi)
\bar F_{12}(\gamma-\i\theta_j-\i\nu_j)
\barbarF_{12}(\gamma-\i\theta_j+\mu-\i\nu_j+\pi),
\\
S_{+-}(\gamma,\mu;\ve,\theta,\nu)
&=\prod_j F_{12}(\gamma-\i\theta_j+\pi)F_{12}(\gamma-\i\theta)
\bar F_{12}(\gamma-\i\theta_j+\mu)
\\
&\quad\times
\prod_{j\in N_-}\bar F_{12}(\gamma-\i\theta_j-\i\nu_j+\pi)
\bar F_{12}(\gamma-\i\theta_j-\i\nu_j)
\barbarF_{12}(\gamma-\i\theta_j+\mu-\i\nu_j),
}\eqlabel\EQFthree
$$
and it is equal to 1 up to $O(\epsilon)$. But the ratio can be estimated
much more precisely by use of Eq.~(\EQratio):
$$
\eqalign{
{S_{+-}(\gamma,\mu;\ve,\theta,\nu)\over S_{-+}(\gamma,\mu;\ve,\theta,\nu)}
&=\prod_j {\bar F_{12}(\gamma-\i\theta_j+\mu)\over
\bar F_{12}(\gamma-\i\theta_j+\mu+\pi)}
\prod_{j\in N_-}{\barbarF_{12}(\gamma-\i\theta_j+\mu-\i\nu_j)
\over \barbarF_{12}(\gamma-\i\theta_j+\mu-\i\nu_j+\pi)}
=\e^{\i\epsilon\Theta/\pi}+O(q),
\\
\Theta&=\sum_{j\in N_+}\theta_j
-\sum_{j\in N_-}(\theta_j+2\nu_j).
}\eqlabel\EQSratio
$$
Note that $\Theta=\i\pi B'/\sqrt2$
with $B'$ being the contribution of $\tilde V_\ve$ into $B$ from
Eq.~(\EQABdef).

We conclude that the factor $F_{-\alpha\alpha}(\gamma,\mu;\ve,\theta,\nu)$
is independent of $\alpha$ up to $O(\epsilon)$, but we are able to
calculate the ratio $F_{+-}/F_{-+}$ with much better precision, namely
up to $O(q)$. This precision allows us to treat the terms with $n=\pm1$
in Eq.~(\EQcpq), which are of the order $q^{1/2}$.

Now consider the zero mode contribution $H_{-\alpha\alpha}$. From
Eq.~(\EQzfunc) we obtain
$$
\eqalignno{
H_{-\alpha\alpha}^{(i)}(\gamma,\mu;\ve,\theta,\nu)
&=\e^{-\i\epsilon A'-{\epsilon\over2}\left({2\gamma\over\pi}-1\right)}
\e^{-{2\epsilon\mu\over\pi}{\alpha+1\over2}}
\exp\left[4\epsilon
\left({1-\alpha\over4}+{B'\over4\sqrt2}-{\mu\over2\pi}\right)^2
-{\epsilon\over4}\right]
\\
&\quad\times
\theta_{3+i}
\left({1-\alpha\over4}+{B'\over4\sqrt2}-{\mu\over2\pi};
{\i\pi\over4\epsilon}\right)
\bigg/\theta_{3+i}\left({1\over4};{\i\pi\over4\epsilon}\right),
\lnlabel\EQHexplicit}
$$
where $A'$ and $B'$ are contributions of $\tilde V_\ve$ into
the constants $A$ and $B$ from Eq.~(\EQABdef). Collecting
equations (\EQFratio), (\EQSratio), and (\EQHexplicit) we
obtain for the functions in the integrand of Eq.~(\EQintegral)
$$
{H_{+-}F_{+-}\over H_{-+}F_{-+}}
=\theta_{4-i}\left({B'\over4\sqrt2}-{\mu\over2\pi};
{\i\pi\over4\epsilon}\right)\bigg/
\theta_{3+i}\left({B'\over4\sqrt2}-{\mu\over2\pi};
{\i\pi\over4\epsilon}\right)+O(q)
\eqlabel\EQfullratio
$$
[recall that $\theta_4(u)=\theta_3(\pm1/2+u)$]. It means that
the difference $F_{-+}-F_{+-}$ is proportional to the difference
of theta functions in the scaling limit,
$$
\theta_{3+i}(u;\i\pi/4\epsilon)
-\theta_{4-i}(u;\i\pi/4\epsilon)
=(-)^i q^{1/2}\cdot 4\cos 2\pi u +O(q).
\eqlabel\EQthetacos
$$
In our case
$$
u=-{\mu\over2\pi}+{B'\over4\sqrt2}
=-{1\over2\pi}\left(\mu-{\i\over2}\Theta\right).
$$
We see that the contributions of the order $q^0$ [terms with $n=0$ in
Eq.~(\EQcpq)] cancel each other in $\sigma^z$, and all form factors
are comparable and
proportional to $q^{1/2}\sim M^{1/2}$. Once we extracted the factor
of this order, we may only take into account the
terms of the order $\epsilon^0$ in all other multipliers.
We obtain the expression
$$
\eqalignno{
&\langle0|\sigma^z(0)
|\ve_1\theta_1,\ldots,\ve_{2n}\theta_{2n}\rangle_{\rm in}^{(i)}
\\
&\qquad
=(-)^{1-i}2(bM)^{1/2}
\int_{C'_1}{d\mu\over2\pi\i}\int_{C'_2}{d^n\nu\over(2\pi)^n}\,
F_{-+}(\mu)F(\ve,\theta,\nu)
\cos\biggl(\mu-{\i\over2}\Theta\biggr),
\lnlabel\EQresult}
$$
where $F_{-+}(\mu)$ and $F(\ve,\theta,\nu)$ are given by
approximate equations (\EQapprox) and (\EQFtwo).
We see that the integrand nearly splits into two factors, one of
which only depends on $\mu$ and the other one on $\nu_j$. The only
factor depending on both $\mu$ and $\nu_j$ is the `zero mode'
factor $\cos(\mu-{\i\over2}\Theta)$.
Note that the function $F_{-+}(\mu)$
is even. Therefore, we may substitute $\cos(\mu-{\i\over2}\Theta)$
in the integrand
by $\cos\mu\,\cos{\i\over2}\Theta$, and the integral factorizes completely.
Consider the function $F_{-+}(\mu)\cos\mu$. Calculating
it numerically by use of the {\it Mathematica} package I found it
to be a constant:
$$
F_{-+}(\mu)\cos\mu=-1.
$$
Hence the integral in $\mu$ is linearly divergent.
This is the problem mentioned above. Therefore we should
consider large $|\mu|\sim\epsilon^{-1}$. Note, that Eqs.\ (\EQFtof c),
(\EQratio), (\EQFratio), (\EQSratio), and therefore (\EQfullratio)
are valid in the case of large $\mu$. The equation (\EQthetacos) is valid
in the whole region $-\pi^2/2\epsilon<\Im\mu<\pi^2/2\epsilon$
except small vicinities
of the ends. To find approximations of $F_{11}$, $C_-$
in the region $|\mu|\gg1$ one can expand summands in $\epsilon n$ everywhere
except the places where $\epsilon n$ enters with a large coefficient $\mu$,
then one can sum the series in the regions of convergence and continue
analytically the answers. Such consideration shows that the integrands
in all form factors of $\sigma^z$ remain factorable and constant in $\mu$
in nearly the
whole region $-\pi^2/2\epsilon<\Im\mu<\pi^2/2\epsilon$. On the other hand the
integration region in $\mu$ for finite $\epsilon$ must be precisely
$[-\i\pi^2/2\epsilon,\i\pi^2/2\epsilon]$.

So we finally obtain
$$
\eqalignno{
&\langle0|\sigma^z(0)
|\ve_1\theta_1,\ldots,\ve_{2n}\theta_{2n}\rangle_{\rm in}^{(i)}
\\
&\qquad
=(-)^i{2\over\pi}(bM)^{1/2}\log\left(4\over bM\right)
\int_{C'_2}{d^n\nu\over(2\pi)^n}\,
F(\ve,\theta,\nu)\ch{\Theta\over2}
\\
&\qquad
=(-)^i\left(2\over\pi\right)^{n+1}(bM)^{1/2}\log\left(4\over bM\right)
\tilde c_+^n\tilde c_-^n
\int_{C'_2}{d^n\nu\over(2\pi)^n}\,
\ch\biggl({1\over2}\sum_{j\in N_+}\theta_j
-{1\over2}\sum_{j\in N_-}(\theta_j+2\nu_j)\biggr)
\\
&\qquad\quad\times
\prod_{j\in N_-}f_2(\nu)
\prod_{j<j'}f_{22}(\theta_{j'}-\theta_j)
\prod_{j<j',j'\in N_-}\bar f_{22}(\theta_{j'}-\theta_j+\nu_{j'})
\prod_{j<j',j\in N_-}\bar f_{22}(\theta_{j'}-\theta_j-\nu_j)
\\
&\qquad\quad\times
\prod_{j<j'\in N_-}\barbarf_{22}(\theta_{j'}-\theta_j+\nu_{j'}-\nu_j).
\lnlabel\EQfinalresult}
$$
For $n=0$ it matches
the well known Baxter--Kelland formula\refs{\BK}
$$
(-)^i\langle\sigma^z\rangle^{(i)}
={(x^2;x^2)_\infty^2\over(-x^2;x^2)_\infty^2}
\simeq\epsilon^{-1} q^{1/2}(2\pi+O(q)).
$$

Let us interpret the result from the point of view of quantum field
theory. First of all, note that every form factor is proportional to
$M^{1/2}$. It means that the scaling dimension of the field
$\sigma^z(x)$ is $1/2$. There are two linearly independent local fields
of this dimension in the critical sine-Gordon theory,
$\e^{\i\varphi/2}$ and $\e^{-\i\varphi/2}$.
We expect that the field $\sigma^z(x)$ is a linear combination of these
fields. Comparison with the result of Ref.~\Refs{\Lukexp} supports this
conclusion and makes it possible to calculate coefficients
of the combination. The decomposition of $\sigma^z$ corresponds to
the decomposition of $\ch\half\Theta$ into two exponentials in
Eq.~(\EQfinalresult). To make it clear, let us pass from the bosonization
of the lattice model to the bosonization of the field theory.

The `conjugate zero mode' operators
$P'$ and $Q'$ survive
in the scaling limit, but the coefficient at $P'^2$ in $D'^{(i)}$
tends to infinity. So we may only
take into account $P'$. The only trace of $Q'$ lies in the requirement
$\sum_j\alpha_j=0$. Considering form factors of lattice objects like $\sigma^z$
we must accurately extract the leading terms in (\EQcpq), as we have done
above. But for the field theory operators we can consider any
expectation $\langle P'|\ldots|P'\rangle$. For $\epsilon\to0$ it
takes the form
$$
H^{(i)}_N(P')
=(bM/4)^{4P'^2/\pi}
\e^{\i(-1)^i\sqrt2 P'}
\e^{\i\beta_N P'}\ldots\,\e^{\i\beta_1 P'}.
\eqlabel\EQzeromode
$$
It means that the zero mode contribution to the scaling dimension
of the fields is equal to $4P'^2/\pi^2$. The contribution (\EQzeromode)
just provides the factors before exponent field in Lukyanov's formula
for vertex operators corresponding to the field $\e^{\i a\varphi}$
for $P'=\pi a/\sqrt2$. Note that in this case $4P'^2/\pi^2=2a^2$ gives
the correct conformal dimension of the field $\e^{\i a\varphi}$.

Consider the oscillators $a_n$. It is tempting to think that the set
of $a_n$ tends in the scaling limit to a continuous set of operators
$a(\eta)$. But it is not quite correct. To see it, let us consider the
field $F_{12}(\gamma)$. The expression for this correlation function
contains the factor $(-1)^n$ [see Eq.~(\EQtptr)], which is rapidly
oscillating in the scaling limit for finite $\gamma$.
It reflects the fact that
the oscillators forming $V_\pm$ differ from those forming $\tilde V_\pm$
by the factor $(-1)^n$ [see Eqs.~(\EQvvdef), (\EQphidef)]. It means
that the set of $a_n$ splits in two continuous sets of operators
$a(\eta)$ and $b(\eta)$,
$$
[a(\eta),a(\eta')]=[b(\eta),b(\eta')]
=\delta(\eta+\eta'){\sh2\eta\sh\eta\over\eta},
\qquad
[a(\eta),b(\eta')]=0.
\eqlabel\EQabdef
$$
Now let us define the operators $\phi(\gamma)$ and $\tilde\phi(\theta)$
as follows%
\nfootnote{The terms $\sim\log x^3$ in (\EQphidef) do not contribute
traces in this limit. In the exact formulas they can be also omitted together
with the factors $x^{3/4}$ in (\EQvvdef).}
$$
\eqalign{
\phi(\gamma)
&={\sqrt2\over\pi}P'\gamma
-\int_{-\infty}^\infty d\eta\,
{a(\eta)\over\i\sh 2\eta}\e^{-|\eta|/2}\e^{{2\over\pi}\eta\gamma},
\\
\tilde\phi(\theta)
&=-{\i\sqrt2\over\pi}P'\theta
+\int_{-\infty}^\infty d\eta\,
{b(\eta)\over\i\sh 2\eta}\e^{|\eta|/2}\e^{\i{2\over\pi}\eta\theta}.
}\eqlabel\EQphietadef
$$
Finally the definition of the vertex operators is the following
$$
\eqalign{
V_+(\gamma)&=g^{1/2}\bc\e^{\i\phi(\gamma)}\ec,
\\
V_-(\gamma)&=g^{1/2}\int_{C'_1}{d\mu\over2\pi\i}\,
{\pi\over\mu^2-\pi^2/4}
\bc\e^{\i\phi(\gamma)-\i\phi(\gamma+\mu-\pi/2)-\i\phi(\gamma+\mu+\pi/2)}\ec,
\\
\tilde V_+(\theta)
&=\sqrt{2\over\pi}g^{-1/2}\bc\e^{\i\tilde\phi(\theta)}\ec,
\\
\tilde V_-(\theta)
&=-\sqrt{2\over\pi}g^{-1/2}\int_{C'_2}{d\nu\over2\pi}\,
{\pi\over\nu^2+\pi^2/4}
\bc\e^{\i\tilde\phi(\theta)-\i\tilde\phi(\theta+\nu+\i\pi/2)
-\i\tilde\phi(\theta+\nu-\i\pi/2)}\ec
}\eqlabel\EQvvetadef
$$
with
$$
\log g=\int_0^\infty{d\eta\over\eta}{\e^{-\eta}(\ch\eta-1)\over2\ch\eta}
$$
if we assume the regularization rule
$$
\int_0^\infty d\eta\left({a_2\over\eta^2}+{a_1\e^{-\eta}\over\eta}
+f(\eta)\right)
\longrightarrow
\int_0^\infty d\eta\,f(\eta),
\qquad
\int_{C'_1}{d\mu\over2\pi\i}\,{\rm const.}={\pi\over2\epsilon}{\rm const.}
$$
for $f(\eta)$ finite at $\eta\to0$.

The vertex operators $\tilde V_\pm(\theta)$ are Lukyanov's vertex
operators from Ref.~\Refs{\Lukexp}.
The factors before the exponential operators are hidden in the zero mode
$P'$. The operators $V_\pm(\gamma)$ can be considered as the primed operators
from Ref.~\Refs{\Lukbos} at the infinitely distant point. They only interact
with the operators $\tilde V_\pm(\theta)$ through the zero mode.
Their physical meaning in the framework of the quantum field theory is vague.
As we can see, the initial idea by Lukyanov that the primed vertex operators
describe local fields was in the right direction, but it is necessary to move
them to the point $\pi^2/2\epsilon$
which is infinitely far for $\epsilon\to0$.

Comparison with Ref.~\Refs{\Lukexp} gives
$$
\sigma^z(x)\sim\cos\half\varphi(x).
\eqlabel\EQcos
$$
Unfortunately we are unable to find the proportionality coefficient
from the vacuum expectations, found by Lukyanov in Ref.~\Refs{\LZ}.
In the limit $\beta\to1$ these expectation values become infinite.
It means that the critical sine-Gordon is not described by a free
boson in the ultraviolet region, and the normalization of the
operators of Ref.~\Refs{\LZ}, based on the conformal field theory,
is not applicable in this case. In the lattice theory it is reflected by
the additional factor $\epsilon^{-1}\sim\log(4/bM)$ in the form
factors. This factor means that the normalization changes with the
scale. Maybe it witnesses about logarithmic factors in the
correlation functions at short distances.

Our last remark concerns the fields\refs{\JM,\Bougourzi}
$\sigma^\pm=\half(\sigma^x\pm\i\sigma^y)$. The leading zero mode contribution
to these fields comes from $n=0$ in (\EQcpq), but the
scaling dimension $1/2$ comes from the oscillators, because the requirement
$\sum_j\alpha_j=0$ gives here $\#N_+ - \#N_- =\pm2$, and the factors depending
on $q$ in (\EQFtof) and  (\EQCtoc) do not cancel. It is easy to understand
from the form factors axioms\refs{\Lukbos} that
the fields $\sigma^\pm(x)$ are semilocal with respect to $\sigma^z(x)$.
There are two semilocal fields in the critical sine-Gordon theory
$\e^{\pm\i\tilde\varphi/2}$ with
$\tilde\varphi(x)=\int_{-\infty}^{x^1}dy^1\,\d_0\varphi(y)$.
So we can suppose that
$$
\sigma^\pm(x)\sim\e^{\pm{\i\over2}\tilde\varphi(x)}.
$$

\nsec Discussion

We conclude that the explicit form of the form factors in
the scaling limit of the six vertex model is consistent with
Lukyanov's conjecture on the free field representation of the form
factors of the exponential fields in the sine-Gordon model
for $\beta=1$, $a=1/2$. A plausible argumentation extends this
consistency to arbitrary values of $a$. This investigation can be
continued in two ways. First, it would be interesting
to calculate form factors
of more general local fields, {\it e.~g.} products of adjacent
polarizations. Second, it is important to generalize this derivation
to other models. For example, the bosonization of the
Andrews--Baxter--Forrester model\refs{\LP} must give in the scaling
limit the bosonization for the restricted sine-Gordon model.

We expect that the divergency in the integral in $\mu$ is
a peculiarity of the six vertex model or of the critical sine-Gordon model.
It is related to the logarithmic factors in the scaling behavior
of quantities, and originates in the dynamical mass generation phenomenon.


\sec* Acknowledgments

I am grateful to R.~Schrader for his hospitality at Freie Universit\"at Berlin
where this work was started under the support by the program
Volkswagen--Stiftung. I am indebted to B.~Davies,
B.~Feigin, S.~Khoroshkin, D.~Lebedev, S.~Lukyanov, S.~Pakuliak, Ya.~Pugai, and
J.~R.~Reyes Mart\'{\char"10}nez for fruitful discussions, and,
especially, to P.~Grinevich for his consultations on elliptic functions.
This work was supported by Civilian Research and Development Foundation
under the grant RP1-277, and by INTAS and RFBR under the grant 95-0690.

\bigskip\allowbreak\bigskip\immediate\closeout\rfile
\vbox{\secfont\noindent References\bigskip}\nobreak
\catcode`@=11\input refs.tmp\catcode`@=12\bigskip

\end